\documentclass[%
reprint,
aps,
pra
]{revtex4-2}

\usepackage[usenames]{color}
\usepackage{colortbl}
\usepackage{graphicx}
\usepackage{dcolumn}
\usepackage{bm}
\usepackage{amsmath}
\usepackage{amssymb}
\usepackage[utf8]{inputenc}

\usepackage{comment}
\usepackage{dsfont}
\usepackage{xcolor}

\begin{document}

\preprint{APS/123-QED}

\title{
Measurement-based acceleration of optical computations 
}

\author{I. V. Vovchenko$^{1,3}$, A. A. Zyablovsky$^{1,2,3}$, A. A. Pukhov$^{1,2}$, E. S. Andrianov$^{1,2,3}$}
\affiliation{%
 $^1$Moscow Center for Advanced Studies (MCAS), 20 Kulakova, Moscow 101000, Russia;
}%
\affiliation{
 $^2$Institute for Theoretical and Applied Electrodynamics, 13 Izhorskaya, Moscow 125412, Russia;
}
\affiliation{%
 $^3$Dukhov Research Institute of Automatics (VNIIA), 22 Sushchevskaya, Moscow 127055, Russia;
}%


\date{\today}

\begin{abstract}
Analog coprocessors are intensively developing nowadays with the aim to optimize energy computations of neural networks. 
In this work we focus on the possibility of using detection of collective oscillations in optical systems for computational purposes.
We show that in a system of coupled resonators, collective oscillations can be used to implement matrix-vector multiplication.
The matrix is formed by the coupling constants between the resonators, and the input vector is formed by the initial occupancies of the involved modes.
The frequency of the collective oscillations is growing with the number of the involved modes, similarly to Rabi oscillations.
The time needed for their detection, i.e., averaging, decreases with an increase in the input vector dimension.
We discuss the limitations imposed on parallel computation in the system by restriction of the allowed optical frequency band.
\end{abstract}

\maketitle


\section{Introduction}
Analog coprocessors~\cite{shastri2021photonics,destras2023survey,ferreira2017progress,ohno2022si,yi2021multi,shastri2021photonics,youngblood2022coherent,giamougiannis2023coherent,kovaios2025chip,pierangeli2021photonic,miscuglio2020photonic,hu2024diffractive,lin2018all,harris2018linear,bogaerts2020programmable,carolan2015universal,
aifer2024thermodynamic,melanson2025thermodynamic,abbas2021power,jia2019quantum,killoran2019continuous,kim2012functional,li2021hardware,9932877,xia2019memristive,ma2017darwin} attract great attention in the field of applied science and are intensively developing nowadays.
They are used in neural networks for acceleration of computations similarly to digital coprocessors~\cite{wu2020energy,lazaro2007hardware,reuther2020survey}.
Digital coprocessors are more precise than analog coprocessors~\cite{jawandhiya2018hardware,dias2003commercial}, while analog coprocessors are faster and more energy efficient~\cite{shastri2021photonics,destras2023survey,ferreira2017progress,li2021hardware,9932877,xia2019memristive,aifer2024thermodynamic,melanson2025thermodynamic,harris2018linear}.
Today, energy efficiency is considered an indispensable feature of a computational system~\cite{shastri2021photonics,destras2023survey,harris2018linear}.
This happens due to the extensive growth of data centers and their electricity consumption, which is evaluated to become of the order of electricity consumption of large countries in the following decade~\cite{ferreira2017progress,xia2019memristive,shehabi20242024,IEA}.

Recently, optical coprocessors implementing linear operations in optical crossbar structures based either on waveguides and microring resonators~\cite{ohno2022si,yi2021multi,shastri2021photonics} or meshes of interferometers~\cite{youngblood2022coherent,giamougiannis2023coherent,shastri2021photonics} are actively investigated as candidates for energy efficiency improvement.
Another approach is to use bosonic modes of free space to perform linear operations~\cite{pierangeli2021photonic,hu2024diffractive,lin2018all}.
The width of the optical frequency band can be involved in computations to operate with tensors of higher dimension~\cite{miscuglio2020photonic,youngblood2022coherent,destras2023survey}.

Important elements of the optical coprocessor are light detectors~\cite{nabet2023photodetectors,rieke2003detection,gundacker2013sipm,seifert2012comprehensive,carmichael2009open}.
They deliver the result achieved in the optical computing scheme to the post-processing stage.
The detectors are also used as non-linear elements to implement electro-optical activation functions~\cite{shastri2021photonics,destras2023survey}.
However, being non-linear elements, the detectors are not directly involved in the implementation of linear operations.

In turn, in the quantum computing measurement process, i.e., detection, is a principal building block of measurement-based quantum computers, which exploit one-way quantum processing of cluster states~\cite{raussendorf2001one,briegel2009measurement,prevedel2007high,kues2019quantum,reimer2019high,wang2020qudits}.
One can ask, could optical computing benefit from a possibility to use detectors not only for detection of the result of calculation but also as a part of calculation processes?

In quantum computations, detectors perform quantum-mechanical averaging of certain operators~\cite{carmichael2009open}.
In classical optics, detectors average oscillations of light field~\cite{nabet2023photodetectors,rieke2003detection,carmichael2009open}. 
In turn, in systems consisting of many coupled modes, collective oscillations appear.
An example of well-studied collective oscillations are Rabi oscillations appearing in systems of atoms coupled to light field modes~\cite{scully1997quantum,barnett2002methods,rahimi2020polariton,lisyansky2024quantum}.
The frequency of Rabi oscillations is growing with the number of involved atoms and photons in cavity~\cite{scully1997quantum,barnett2002methods,rahimi2020polariton,fregoni2022theoretical}.
Then, it is expected that the frequencies of the collective oscillations appearing in the systems of many coupled modes should be similarly growing with the number of the involved modes.
If so, the time that is required for detection, i.e., averaging, of such collective oscillations will decrease with an increase in the number of the involved modes.
This can increase the speed of optical computing.

In this work we propose measurement-based acceleration of an optical coprocessor consisting of several resonators (the side resonators) coupled to a general one (the central resonator).
We show that collective Rabi-like oscillations appear in the system.
Being averaged, i.e., measured, occupancies of the central resonator's modes become proportional to the weighted initial occupancies of the side resonators' modes.
We show that this can be used to implement matrix-vector multiplication.
The matrix is formed by the coupling constants between the resonators' modes.
The input vector is formed by the initial occupancies of the resonators' modes.
The frequency of the collective Rabi-like oscillations is growing as a square root of the number of the involved modes.
Hence, the time needed for the averaging, i.e., the computational time of the mentioned linear operation, is respectively decreasing with an increase in the input vector dimension.
We discuss the limitations imposed on parallel computations by the restriction of the used optical frequency band.
We show that the rate of parallel computations is primarily limited by the amount of cross-talk between modes of the side and central resonators that are non-resonant.
We estimate the computational rate per one optical frequency as $10^{13}$ Hz.

\section{The model}
\begin{figure}
    \centering
    \includegraphics[width=0.75\linewidth]{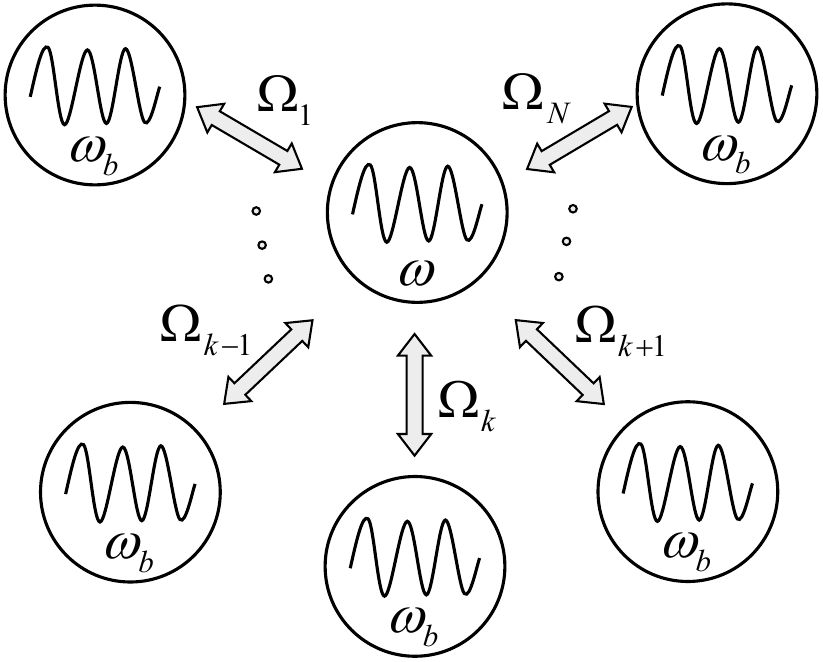}
    \caption{Schematic representation of the system with Hamiltonian Eq.~(\ref{Ham}).}
    \label{System1_img}
\end{figure}

To describe the dynamics of $N$ side resonators coupled to the central one,
let consider the simplified model of one-mode resonators first, see Fig.~\ref{System1_img}.
Further, we address the modes as side and central, respectively.
The Hamiltonian of the system in the rotating wave approximation equals~\cite{carmichael2009open,lisyansky2024quantum,breuer2002theory}
\begin{equation}\label{Ham}
    \hat{H}=\omega\hat{a}^\dag \hat{a}+\sum_{k=1}^N \omega_b \hat{b}^\dag_k \hat{b}_k + \sum_{k=1}^N \Omega_k(\hat{a}^\dag\hat{b}_k +\hat{b}_k^\dag \hat{a}).
\end{equation}
Here $\omega$ is the frequency of the central mode, $\omega_b$ is the frequency of the side modes, $\Omega_k$ is the coupling constant between the $k$-th side mode and the central mode, $\hbar=1$.
Using Heisenberg representation, we get
\begin{gather}
    \dot{\hat{a}}(t)=-i\omega \hat{a}(t) - i \sum_{k=1}^N \Omega_k \hat{b}_k(t),
    \\ \nonumber
    \dot{\hat{b}}_k(t)=-i\omega_b \hat{b}_k(t) - i\Omega_k \hat{a}(t).
\end{gather}

Introducing operator $\hat b(t)=\sum\limits_{k=1}^N\Omega_k\hat{b}_k(t)/\sum\limits_{k=1}^N \Omega_k^2$, one can see that the dynamics of $\hat{a}(t)$ depends only on $\hat{b}(t)$, and the dynamics of $\hat{b}(t)$ depends only on $\hat{a}(t)$.
Solving this set of two differential equations, one gets
\begin{gather}
    \hat{a}(t)=\left(\cos{\Omega_\mathrm{R} t - \frac{i\Delta/2}{\Omega_\mathrm{R}} \sin{\Omega_\mathrm{R} t}}\right)  \hat{a}(0) e^{-i\frac{\omega+\omega_b}{2}t} \\ \nonumber
    - i \sin{\Omega_\mathrm{R} t} \left(\frac{\vec{\Omega}}{\Omega_\mathrm{R}},\vec{\hat{b}}_0\right)e^{-i\frac{\omega+\omega_b}{2}t},
\end{gather}
Here
$\vec{\hat{b}}_0=(\hat{b}_1 (0),\ldots, \hat{b}_N (0))^T$,
$\vec{\Omega}=(\Omega_1,\ldots,\Omega_N)^T$,
detuning is denoted as $\Delta=\omega-\omega_b$,
$\Omega_\mathrm{R}=\sqrt{\left(\frac{\Delta}{2}\right)^2+(\vec{\Omega},\vec{\Omega})}$,
$(\cdot,\cdot)$ denotes the usual Euclidean dot product.
Thus,
\begin{gather}
    \hat{a}^\dag (t) \hat{a}(t) = \left(\cos^2{\Omega_\mathrm{R} t }+ \frac{(\Delta/2)^2}{\Omega_\mathrm{R}^2} \sin^2{\Omega_\mathrm{R} t}\right)  \hat{a}^\dag_0\hat{a}_0 \\ \nonumber
    +\sin^2{\Omega_\mathrm{R}t}\sum_{k=1}^N \frac{\Omega_k^2}{\Omega_\mathrm{R}^2} \hat{b}^\dag_{k0} \hat{b}_{k0} 
    +\sin^2{\Omega_\mathrm{R}t}
    \sum_{k\ne q=1}^N \frac{\Omega_k \Omega_q}{\Omega_\mathrm{R}^2}\hat{b}_{k0}^\dag \hat{b}_{q0}\\ \nonumber
    -i\left(\cos{\Omega_\mathrm{R} t }+ \frac{i\Delta/2}{\Omega_\mathrm{R}} \sin{\Omega_\mathrm{R} t}\right) \sin{\Omega_\mathrm{R} t}\, \sum_{k=1}^N \frac{\Omega_k}{\Omega_\mathrm{R}} \hat{a}^\dag_0 \hat{b}_{k0} \\ \nonumber
    +i\left(\cos{\Omega_\mathrm{R} t} - \frac{i\Delta/2}{\Omega_\mathrm{R}} \sin{\Omega_\mathrm{R} t}\right) \sin{\Omega_\mathrm{R} t}\, \sum_{k=1}^N\frac{\Omega_k}{\Omega_\mathrm{R}}  \hat{b}^\dag_{k0} \hat{a}_0.
\end{gather}
Here, $\hat{a}_0=\hat{a}(0)$, $\hat{b}_{k0}=\hat{b}_k (0)$.

It is seen that collective oscillation at the frequency $\Omega_\mathrm{R}$ influences the occupancy $\hat{a}^\dag (t) \hat{a}(t)$.
If $\Delta=0$, $\Omega_\mathrm{R}\sim\sqrt{N}$.
Thus, the frequency of the collective oscillation $\Omega_\mathrm{R}$ grows with an increase in the number of the modes involved in these collective oscillations, similarly to the way the frequency of Rabi oscillations grows with an increase in the number of involved atoms or photons in cavity~\cite{scully1997quantum,barnett2002methods,rahimi2020polariton}.
In view of this, further, we address this collective oscillation also as Rabi-like.

Detection of light averages its intensity over a time interval~\cite{nabet2023photodetectors,rieke2003detection,gundacker2013sipm,seifert2012comprehensive,carmichael2009open}.
We can introduce the following function of time
$
\overline{\hat{a}^\dag (t) \hat{a}(t)}_{t} 
\equiv 
\frac{1}{t}\int_0^{t} \hat{a}^\dag (t') \hat{a}(t') dt'$
to describe the averaging of the central mode intensity.
If $t=t_0\gg 1/\Omega_\mathrm{R}$, we get 
\begin{gather}\label{AvA}
    \overline{\hat{a}^\dag (t) \hat{a}(t)}_{t_0} 
    =
    \left(1 -\frac{(\vec{\Omega},\vec{\Omega})}{2\Omega_\mathrm{R}^2} \right)  \hat{a}^\dag_0\hat{a}_0 +\sum_{k=1}^N \frac{\Omega_k^2}{2\Omega_\mathrm{R}^2} \hat{b}^\dag_{k0} \hat{b}_{k0} 
    \\ \nonumber
    +\sum_{k=1}^N\frac{\Delta \Omega_k}{2\Omega_\mathrm{R}^2}(\hat{a}^\dag_0\hat{b}_{k0} +\hat{b}^\dag_{k0}\hat{a}_0)
    +\sum_{k\ne q=1}^N \frac{\Omega_k \Omega_q}{2\Omega_\mathrm{R}^2}\hat{b}_{k0}^\dag \hat{b}_{q0}.
\end{gather}

Eq.~(\ref{AvA}) can be rewritten as
\begin{gather}
\nonumber
    \overline{\hat{a}^\dag (t) \hat{a}(t)}_{t_0}  
    =
    \hat{a}^\dag_0\hat{a}_0+\frac{(\vec{\Omega},\vec{\Omega})}{2\Omega_\mathrm{R}^2} \left(\sum_{k=1}^N \frac{\Omega_k^2}{(\vec{\Omega},\vec{\Omega})} \hat{b}^\dag_{k0} \hat{b}_{k0}-\hat{a}^\dag_0 \hat{a}_0\right)  
    \\
    +\sum_{k=1}^N\frac{\Delta \Omega_k}{2\Omega_\mathrm{R}^2}(\hat{a}^\dag_0\hat{b}_{k0} +\hat{b}^\dag_{k0}\hat{a}_0)
    +\sum_{k\ne q=1}^N \frac{\Omega_k \Omega_q}{2\Omega_\mathrm{R}^2}\hat{b}_{k0}^\dag \hat{b}_{q0}.
\end{gather}

Then, occupancy of the central mode averaged over time interval $t_0$ equals
\begin{gather}
\nonumber
    \overline{\langle\hat{a}^\dag \hat{a}\rangle}_{t_0}  =
    \langle\hat{a}^\dag\hat{a}\rangle_0+\frac{(\vec{\Omega},\vec{\Omega})}{2\Omega_\mathrm{R}^2} \left(\sum_{k=1}^N \frac{\Omega_k^2}{(\vec{\Omega},\vec{\Omega})} \langle\hat{b}^\dag_{k} \hat{b}_{k} \rangle_0-\langle\hat{a}^\dag \hat{a}\rangle_0\right)  
    \\
    +\sum_{k=1}^N\frac{\Omega_k \Delta}{2\Omega_\mathrm{R}^2}\left(\langle\hat{a}^\dag\hat{b}_{k}\rangle_0 + \langle\hat{b}^\dag_{k}\hat{a}\rangle_0\right)
    +\sum_{k\ne q=1}^N \frac{\Omega_k \Omega_q}{2\Omega_\mathrm{R}^2}\langle\hat{b}_{k}^\dag \hat{b}_{q}\rangle_0.
\end{gather}
Here, as previously for operators, index $0$ denotes the value of the occupancy at $t=0$.
Let $\langle\hat{a}^\dag\hat{a}\rangle_0=0$, ${\rm Re}\langle \hat{a}^\dag\hat{b}_k\rangle_0=0$, and ${\rm Re}\langle\hat{b}_{k}^\dag\hat{b}_{q}\rangle_0\sim \delta_{kq}$, then
\begin{gather}\label{nnn}
    \overline{\langle\hat{a}^\dag \hat{a}\rangle}_{t_0} 
    =\frac{(\vec{\Omega},\vec{\Omega})}{2\Omega_\mathrm{R}^2} \sum_{k=1}^N \frac{\Omega_k^2}{(\vec{\Omega},\vec{\Omega})} \langle\hat{b}^\dag_{k} \hat{b}_{k} \rangle_0
    \equiv\frac{(\vec{\Omega},\vec{\Omega})}{2\Omega_\mathrm{R}^2} \tilde{n}_b.
\end{gather}
Here $\tilde{n}_b=\sum_{k=1}^N {\Omega_k^2} \langle\hat{b}^\dag_{k} \hat{b}_{k} \rangle_0/(\vec{\Omega},\vec{\Omega})$ is weighted initial occupancy of side modes.
The weight coefficients are equal $\Omega_k^2/(\vec{\Omega},\vec{\Omega})$.

\begin{figure}
    \centering
    \includegraphics[width=0.9\linewidth]{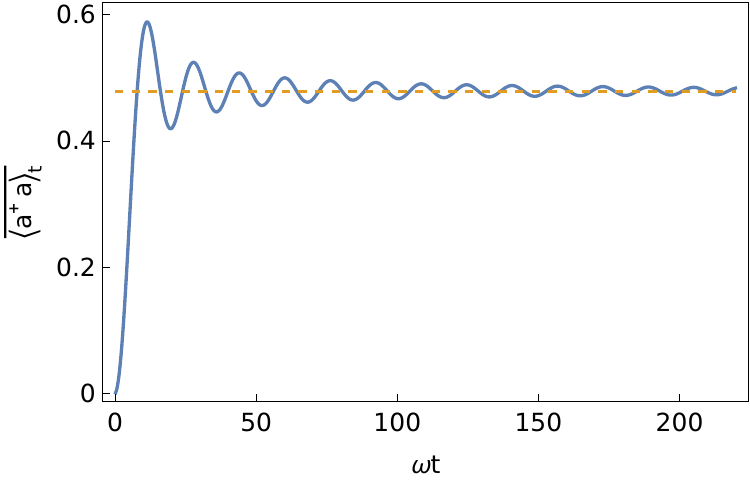}
    \caption{The dynamics of $\overline{\langle\hat{a}^\dag\hat{a}\rangle}_t$ (blue solid line) and $0.5\tilde n_b$ (orange dashed line).
    The parameters: $\omega=\omega_b=1$, $N=50$, $\Omega_k$ are set randomly between $5\cdot 10^{-3}$ and $5\cdot 10^{-2}$. 
    Initial occupancies $\langle\hat{b}_k^\dag \hat{b}_k\rangle_0$ are set randomly between $0.1$ and $1$, $\langle\hat{a}^\dag\hat{a}\rangle_0=0$.}
    \label{aa_sr}
\end{figure}

\section{Implementation of matrix-vector multiplication}

\begin{figure}
    \centering
    \includegraphics[width=0.9\linewidth]{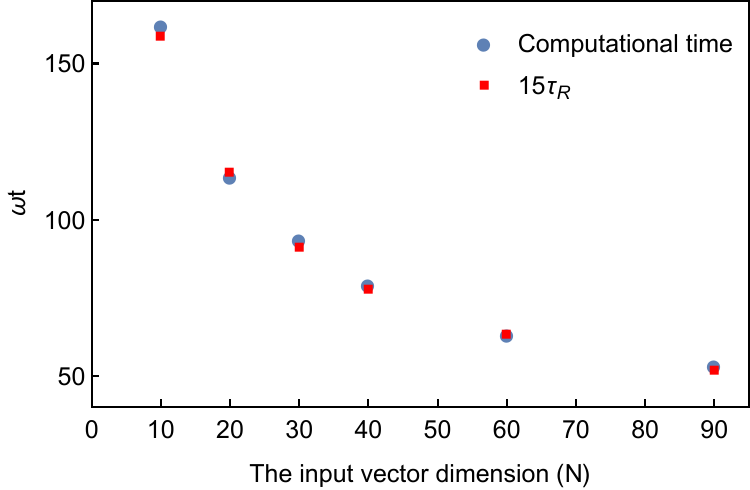}
    \caption{The average value of the computational time providing $3\%$ precision (blue circles) and $15\tau_R$ (red squares).
    The parameters are set similarly to Fig.~\ref{aa_sr}.}
    \label{Comp_time}
\end{figure}

Let $\Delta=0$, then $\overline{\langle\hat{a}^\dag \hat{a}\rangle}_{t_0} = 0.5\tilde{n}_b$.
The dynamics of $\overline{\langle\hat{a}^\dag \hat{a}\rangle}_t$ is presented in Fig.~\ref{aa_sr}.
It is seen that $\overline{\langle\hat{a}^\dag \hat{a}\rangle}_t$ indeed tends to the value $0.5 \tilde n_b$ while $t\rightarrow +\infty$.
This requires $t_0\approx10\tau_R$, where $\tau_R=\Omega_R^{-1}$ (in Fig.~\ref{Comp_time} it is shown that for $3\%$ precision of the convergence, the time $15\tau_R$ is required).

This dynamics of $\overline{\langle\hat{a}^\dag \hat{a}\rangle}_t $ can be used for implementation of dot product.
Indeed, let 
\begin{gather}
     \vec{A}=(\Omega_1^2,\ldots,\Omega_N^2)^T/(\vec{\Omega},\vec{\Omega}),\\ \nonumber
     \vec{B}=(\langle\hat{b}^\dag_{1} \hat{b}_{1} \rangle_0,\ldots,\langle\hat{b}^\dag_{N} \hat{b}_{N} \rangle_0)^T, 
\\ \nonumber
\text{ then }\overline{\langle\hat{a}^\dag \hat{a}\rangle}_{t_0}=0.5(\vec{A},\vec{B}).
\end{gather}
Thus, to implement the dot product of two positive vectors (one of which is normalized) in this system, one should tune normalized squares of coupling constants to form the first vector and excite the side modes occupancies to form the second vector.
Then, the result of the dot product operation would be stored in the averaged value of the central mode occupancy.

If one gets several systems with Hamiltonian Eq.~(\ref{Ham}), he can implement multiple dot products in parallel or matrix-vector multiplication.
(for the last one, occupancies of the side modes should be the same in different systems).
With this scope, one can replicate the system or use the frequency domain of the resonators.
Here we consider the second option.

Let consider $N$ side resonators coupled to the central one.
All resonators are similar and have $M$ different modes.
Let denote the frequency step in the resonators as $\Delta_b$.
The model of one-mode resonators can be applied to this case too.
For that, $\Delta_b$ should be sufficient to neglect both linear and non-linear cross-talk between non-resonant modes of side resonators via their coupling to the central resonator.
If so, the model from the previous section can be applied to each mode of the central resonator independently.
Hence, one can compute $M$ independent dot products using this system.
If the occupancies of the involved side resonators' modes are equal, one can realize matrix-vector multiplication
\begin{gather}
\nonumber
    \left[
    \begin{array}{cc}
        \overline{\langle\hat{a}^\dag \hat{a}\rangle}_{t1}\\
        \vdots\\
        \overline{\langle\hat{a}^\dag \hat{a}\rangle}_{tM}
    \end{array}
    \right]
    \approx
    \frac{1}{2}
    \left[
    \begin{array}{ccc}
        p_{11} & \cdots  & p_{1N}\\
        \vdots & \ddots  & \vdots \\
        p_{M1} & \cdots & p_{MN}
    \end{array}
    \right]
    \left[
    \begin{array}{cc}
        \langle\hat{b}^\dag_{1} \hat{b}_{1} \rangle_0\\
        \vdots\\
        \langle\hat{b}^\dag_{N} \hat{b}_{N} \rangle_0
    \end{array}
    \right],
    \\ 
    \text{where }
    p_{mk}
    =
    \left(\Omega_{k}^{(m)}\right)^2/{\sum\limits_{q=1}^N\left(\Omega_{q}^{(m)}\right)^2}.
\end{gather}
Here $\overline{\langle\hat{a}^\dag \hat{a}\rangle}_{tm}$ is the averaged occupancy of the central resonator's mode resonant to the $m$-th mode of the side resonators, $1\le m\le M$, $\Omega_k^{(m)}$ is the coupling constant between the $m$-th mode of the $k$-th side resonator and the central resonator's mode that is resonant to this mode of the side resonator.
It is seen that the matrix involved in the computations is a stochastic matrix, i.e., a transition matrix (kernel) of a Markov process (Markov chain)~\cite{norris1998markov,douc2018markov,freedman2012markov,agbinya2022applied,tweedie2001markov}.
Stochastic matrices are used in many sciences~\cite{tweedie2001markov,brooks1998markov,tamir1998applications,agbinya2022applied,sigaud2013markov,torra2021space}, including artificial intelligence and neural networks~\cite{agbinya2022applied,sigaud2013markov,torra2021space,pardo2025neural}.

Notably, the computational time that is of the order of $\tau_R$ decreases as $1/\sqrt{N}$ with an increase in the input vector dimension.
Indeed, see Fig.~\ref{Comp_time}.
This happens due to the growth in frequency of the collective Rabi-like oscillations with an increase in $N$.
This leads to the decrease in the time needed for the averaging, i.e., the computational time.

Also note that the factor $(\vec{\Omega},\vec{\Omega})/2\Omega_\mathrm{R}^2$ can be used to discriminate some of the inputs and outputs.
With proper feedback loop design, this can help implementing the learning process used in neural networks~\cite{abdi1999neural,gurney2018introduction,islam2019overview,muller2012neural}.

\section{Limitation imposed by geometry and restricted frequency band}

The frequency band that can be used in computations is usually restricted by the frequency response of the detecting system~\cite{nabet2023photodetectors,rieke2003detection,gundacker2013sipm,seifert2012comprehensive}.
For small matrices the influence of the finiteness of the frequency band can be neglected, as only several optical frequencies are involved in computations.
Maximization of computational rate requires the usage of the whole frequency band and maximization of the number of computations per optical frequency.
However, the maximization of computations per optical frequency leads to more intense non-resonant cross-talk between modes (see previous Section).
This restricts the number of optical frequencies that can be used in parallel computations.

Let the frequency band that is used in computations be restricted by the width $\delta$.
Let $\Omega_{max}^{(\Delta)}$ be the maximal value of the non-resonant cross-talk coupling constant.
We can use Eq.~(\ref{nnn}) to estimate the needed $\Delta_b$.
Indeed, the influence of non-resonant interaction on the averaged occupancy of the central resonator's modes decreases as $\Delta^{-2}$ with increase in detuning $\Delta$, i.e., $(\vec{\Omega},\vec{\Omega})/2\Omega_\mathrm{R}^2=4(\vec{\Omega},\vec{\Omega})/(\Delta^2+4(\vec{\Omega},\vec{\Omega}))$.
The $(\vec{\Omega},\vec{\Omega})$ can be seen as the width of the Lorenz line.
Then, to neglect the effect of cross-talk between non-resonant modes, the frequency step in resonators $\Delta_b$ should satisfy $\Delta_b\gg \sqrt{N}\Omega_{max}^{(\Delta)}$.

Hence, the maximal number of the optical frequencies that can be used in the computations equals $M=\delta/\sqrt{N}\Omega_{max}^{(\Delta)}$.
Thus, the system can implement $N\delta/\sqrt{N}\Omega_{max}^{(\Delta)}$ elementary operations at once.
The required time equals $\tau\sim\Omega_{\rm R}^{-1}\sim N^{-1/2} \Omega_{max}^{-1}$, where $\Omega_{max}=\max\limits_{k,m}\Omega_k^{(m)}$.
Let introduce parameter $\alpha=\Omega_{max}^{(\Delta)}/\Omega_{max}$
which describes the ratio of non-resonant coupling to the resonant one.
Then the rate of computations (number of elementary operations performed per second) can be estimated as $NM/\tau\sim\alpha^{-1} N \delta$.

One can see that the minimal rate of computations equals $N\delta$ (obviously $\alpha \le 1$).
It does not depend on coupling constants between the resonators.
It increases linearly with an increase in the number of entries per optical frequency $N$ and is inversely proportional to $\alpha$.
Thus, for fixed $N$, the maximization of the computational rate in a restricted frequency band is a problem of minimization of the cross-talk between non-resonant modes, i.e., minimization of $\alpha$.

In turn, the maximal value of $N$ is restricted in the used model.
For the described model to be applicable, the time that is needed for the light to travel around the central resonator should be much less than the period of the collective Rabi-like oscillation.
Then, $Nl_{max}/c\ll1/\Omega_{max}\sqrt{N}$, i.e., $N\ll(c/\Omega_{max}l_{max})^{2/3}$, where $l_{max}$ is the maximal coupling length between a side resonator and the central resonator.
Let $l_{max}\sim 10^{-6}$ m, $\Omega_{max}\sim10^{9}$ s$^{-1}$ (it should be greater than the dissipation rate, which is $\sim10^{7}\div 10^{8}$ s$^{-1}$ in resonators with high $Q$-factor~\cite{wu2020greater,shitikov2018billion,kim2020universal}), then $N\ll 10^4$.
Thus, at a single optical frequency, we get $N/\tau\ll N^{3/2}\Omega_{max}\ll c/l_{max}\sim 10^{14}$ Hz computational rate.
Hence, the maximum value of the computational rate per optical frequency can be estimated as $10^{13}$ Hz.

Let the optical frequency band restricted by $1.5$ $\mu$m and $0.4$ $\mu$m be involved in the computations.
Then, as for $\alpha\sim 1$ maximal value of $M\sim\delta/\sqrt{N}\Omega_{max}$, we can expect about $1000$ optical frequencies computing in parallel, and the computational rate $N\delta\sim 10^{16}$ Hz in the system.
Thus, the employment of parallel computations can increase the computational rate in the system by three orders of magnitude.

\section{Conclusion and discussion}

In this work we describe measurement-based acceleration of an optical coprocessor.
The considered coprocessor is based on side resonators coupled to a central resonator.
We show that if the frequency step in the resonators is sufficient to neglect cross-talk between non-resonant modes, the averaged occupancies of the central resonator's modes are proportional to the weighted initial occupancies of the side resonators' modes.
We show that this can be used to implement matrix-vector multiplications.
The input vector is encoded in initial occupancies of the resonators' modes.
The matrix is encoded by squares of coupling constants between the resonators' modes.
The output vector is encoded in the averaged occupancies of the central resonator's modes.

The averaging, i.e., measurement, conducted by a detector levels out the collective Rabi-like oscillations that appear in the system.
The frequency of the Rabi-like oscillations is growing as a square root of the number of the involved modes.
Hence, the time that is needed for the averaging, i.e., the computational time, is decreasing as an inverse square root with an increase in the input vector dimension.

Finally, we discuss the limitations imposed on the parallel computations in the optical frequency band by the restriction of the optical frequency band that can be used.
We show that it is impossible to maximize both the number of simultaneous computations per optical frequency and the number of the optical frequencies in parallel computations at the same time.
We demonstrate that this happens due to cross-talk between non-resonant modes of the resonators.
Thus, the computational rate in a restricted optical frequency band is proportional to the number of simultaneous computations per optical frequency and to the width of the used frequency band, and it is inversely proportional to the value of the non-resonant cross-talk.
The computational rate per optical frequency can be estimated as $10^{13}$ Hz.

Note, we discuss the effect of collective Rabi-like oscillation on computations in an optical coprocessor.
However, similar effects can be found in other frequency bands (microwave, for example) and systems of quasiparticles obeying quadratic Hamiltonians, i.e., phonons~\cite{minarik2013hamiltonian,zaitsev2008introduction,scherer1984theory}, magnons~\cite{yuan2017magnon,Kittel1991quantum,mcclarty2022topological}, and plasmons~\cite{tame2013quantum,downing2017topological,biehs2013dynamical}.

\textbf{Acknowledgment.}
A.A.Z. and E.S.A. acknowledge the support of the Foundation for the Advancement of Theoretical Physics and Mathematics BASIS.

\bibliography{CwCM}

\end{document}